\documentclass[%
 aip,
 amsmath,amssymb,
 reprint,%
]{revtex4-1}

\usepackage{graphicx}
\usepackage{dcolumn}
\usepackage{bm}

\usepackage[utf8]{inputenc}
\usepackage[T1]{fontenc}
\usepackage{mathptmx}
\usepackage{etoolbox}

\usepackage{lineno,comment,siunitx,ulem,xcolor,textgreek}

\makeatletter
\def\@email#1#2{%
 \endgroup
 \patchcmd{\titleblock@produce}
  {\frontmatter@RRAPformat}
  {\frontmatter@RRAPformat{\produce@RRAP{*#1\href{mailto:#2}{#2}}}\frontmatter@RRAPformat}
  {}{}
}%
\makeatother

\begin{document}

\preprint{AIP/123-QED}

\title[Ytterbium-laser-driven THz generation in thin lithium niobate at 1.9 kW average power in a passive enhancement cavity]{Ytterbium-laser-driven THz generation in thin lithium niobate at 1.9 kW average power in a passive enhancement cavity}

\author{Edoardo Suerra}
    \affiliation{\mbox{Dipartimento di Fisica, Università degli Studi di Milano, via Celoria 16, 20133, Milan, Italy}}
    \affiliation{\mbox{Istituto Nazionale di Fisica Nucleare, Sezione di Milano, via Celoria 16, 20133, Milan, Italy}}
\author{Francesco Canella}
    \affiliation{\mbox{Istituto di Fotonica e Nanotecnologie–Consiglio Nazionale delle Ricerche, Piazza Leonardo da Vinci 32, 20133, Milan, Italy}}
    \email{francesco.canella@cnr.it}
\author{Dario Giannotti}
    \affiliation{\mbox{Dipartimento di Fisica, Politecnico di Milano, Piazza Leonardo da Vinci 32, 20133, Milan, Italy}}
    \affiliation{\mbox{Istituto Nazionale di Fisica Nucleare, Sezione di Milano, via Celoria 16, 20133, Milan, Italy}}
\author{Mohsen Khalili}
    \affiliation{\mbox{Photonics and Ultrafast Laser Science (PULS), Ruhr-Universität Bochum, Universitätsstraße 150, 44801, Bochum, Germany}}
\author{Yicheng Wang}
    \affiliation{\mbox{Photonics and Ultrafast Laser Science (PULS), Ruhr-Universität Bochum, Universitätsstraße 150, 44801, Bochum, Germany}}
\author{Kore Hasse}
    \affiliation{\mbox{Experimental Physics and Material Sciences, Helmut-Schmidt-Universität, Holstenhofweg 85, 22043, Hamburg, Germany}}
\author{Sergiy Suntsov}
    \affiliation{\mbox{Experimental Physics and Material Sciences, Helmut-Schmidt-Universität, Holstenhofweg 85, 22043, Hamburg, Germany}}
\author{Detlef Kip}
    \affiliation{\mbox{Experimental Physics and Material Sciences, Helmut-Schmidt-Universität, Holstenhofweg 85, 22043, Hamburg, Germany}}
\author{Clara Saraceno}
    \affiliation{\mbox{Photonics and Ultrafast Laser Science (PULS), Ruhr-Universität Bochum, Universitätsstraße 150, 44801, Bochum, Germany}}
\author{Simone Cialdi}
    \affiliation{\mbox{Dipartimento di Fisica, Università degli Studi di Milano, via Celoria 16, 20133, Milan, Italy}}
    \affiliation{\mbox{Istituto Nazionale di Fisica Nucleare, Sezione di Milano, via Celoria 16, 20133, Milan, Italy}}
\author{Gianluca Galzerano}
    \affiliation{\mbox{Istituto di Fotonica e Nanotecnologie–Consiglio Nazionale delle Ricerche, Piazza Leonardo da Vinci 32, 20133, Milan, Italy}}
    \affiliation{\mbox{Istituto Nazionale di Fisica Nucleare, Sezione di Milano, via Celoria 16, 20133, Milan, Italy}}

\date{\today}

\begin{abstract}
Single-cycle, high-power, high-repetition-rate THz pulse sources are becoming the cornerstone of several scientific and industrial applications.
A promising and versatile method for high-power THz generation is optical rectification in nonlinear crystals pumped by powerful near-infrared ultrafast laser systems.
In this context, ytterbium-based laser sources are particularly advantageous in terms of power scalability and technology establishment.
However, as the repetition rate increases toward hundreds of MHz, the conversion efficiency typically decreases, as most laser systems do not reach sufficiently high average power to correspondingly enhance the peak power to drive the nonlinear conversion process efficiently.
An alternative approach to achieving sufficiently high average power at high repetition rate is based on passive enhancement cavities, which boost the pulse energy of standard watt-level ytterbium lasers by orders of magnitude.
We present the first demonstration of optical rectification in a passive enhancement cavity at multi-kW levels, achieved by a 240-fold power enhancement.
By irradiating a 50-{\textmu}m thin lithium niobate plate with 1.9-kW average power inside the enhancement cavity, we generate milliwatt-level THz pulses with 2-THz bandwidth and 93-MHz repetition rate, mostly limited by the driving pulse duration.
To the best of our knowledge, this represents the highest driving average power used for OR.
This methodology represents a promising new step towards high-repetition-rate and high average power single-cycle THz sources using widely available multi-watt level Yb lasers.
\end{abstract}

\maketitle

\section{\label{sec:intro}Introduction}
Terahertz time-domain spectroscopy (THz-TDS) is nowadays considered a fundamental tool to investigate the millimeter and sub-millimeter regions of the electromagnetic spectrum, finding a variety of applications both in scientific research and in industry.
For instance, THz-TDS is used for hyperspectral imaging of biological samples, non-destructive testing, molecular vibrational spectroscopy, and quantum materials characterization \cite{Koch2023, Neu2018}.
Single-cycle phase-stable THz pulses are highly desirable for TDS, particularly when combined with high repetition rates, ideally exceeding tens of MHz.
In this regime, the signal-to-noise ratio can be enhanced by fast averaging, significantly reducing acquisition times, which is essential for capturing transient phenomena with high temporal resolution \cite{Jepsen2011,Mansourzadeh2021}.
However, despite the great effort of the scientific community, the average power of pulsed THz sources has always been a limiting factor, especially at hundreds of \SI{}{\mega\hertz} repetition rates as pulse energy lacks.
In this context, promising candidates for high-repetition-rate THz generation include photoconductive antennas (PCAs) and sources based on optical rectification (OR) of high-power, ultrafast infrared lasers.
THz emission in a PCA is based on charge carriers generated by an ultrafast laser pulse and accelerated by an external bias field \cite{Burford2017}.
State-of-the-art PCAs can generate THz powers at the milliwatt level, such as \SI{1}{\milli\watt} at \SI{80}{\mega\hertz} with \SI{50}{\milli\watt} of optical power, reaching a dynamic range of \SI{137}{\decibel} in TDS setups \cite{PCA137db}.
Similarly, \SI{4}{\milli\watt} of THz power has been achieved at \SI{78}{\mega\hertz} with \SI{720}{\milli\watt} of pump power using plasmonic structures \cite{Bashirpour2019}.
The limiting factors for the THz power generated by PCAs are typically attributed to the saturation of optical carriers, which constrains the excitation pump power \cite{Burford2017}. This effect becomes more pronounced as the optical pump is focused on a smaller area.
Advanced PCAs with large-area microstructured electrodes (\SI{1}{}–\SI{100}{\milli\meter\squared}) allow higher optical power and show good conversion efficiencies, approximately $2 \times 10^{-3}$, achieving \SI{1.5}{\milli\watt} of THz power at \SI{250}{\kilo\hertz} with \SI{800}{\milli\watt} of pump.
However, at MHz repetition rates, thermal load is a key limitation and only very recent studies suggest mitigation strategies \cite{Khalili2024}.
With the advancement of high-power ytterbium-based ultrafast lasers \cite{Saraceno2019, Muller2020}, infrared laser-driven THz generation via OR in $\chi^{(2)}$ crystals has become a commonly adopted approach to achieve milliwatt-level THz sources.
These lasers enable significant THz power output \cite{Kramer2020, Buldt2021, Vogel2024}, providing high efficiency (up to the percent level), broad bandwidth, and frequency tunability across a range of nonlinear crystals \cite{Fulop2020}.
For instance, the highest average power at MHz repetition rates has been achieved using the tilted pulse front scheme in bulk lithium niobate, which produced \SI{66}{\milli\watt} at \SI{13}{\mega\hertz} with a \SI{2}{\tera\hertz} bandwidth, although this approach involves complex velocity-matching requirements \cite{Meyer2020}.
Another approach using the organic crystal BNA (N-benzyl-2-methyl-4-nitroaniline) achieved \SI{0.950}{\milli\watt} THz power at \SI{13}{\mega\hertz} with a \SI{2.4}{\watt} pump, extending the THz spectrum up to \SI{6}{\tera\hertz} \cite{Mansourzadeh2021b}.
Additionally, organic crystals like HMQ-TMS have demonstrated \SI{1.5}{\milli\watt} output at \SI{10}{\mega\hertz} with a \SI{2.5}{\watt} pump \cite{Buchmann2020}.
While photoconductive emitters are often limited by thermal constraints, nonlinear OR sources, though more complex, offer scalable paths to higher THz power levels needed for demanding applications.
This scalability and versatility position nonlinear methods as vital to advancing high-power THz sources, even as improvements in photoconductive emitter efficiency continue.
On the other hand, OR heavily relies on high driving pulse energies, thus attempts to further increase the repetition rate toward hundreds of MHz result in minuscule conversion efficiencies, even using \sout{large} complex and elaborate systems.  
However, important applications like hyperspectral imaging or molecular spectroscopy take advantage of hundred-\SI{}{\mega\hertz} repetition rates, where producing high THz power is still challenging (see e.g., Refs.~\cite{Weiss2001, Nagai2004}). 
The main limiting factor of using such high repetition rates lies in the high driving energy necessary for significant THz generation via OR, which implies an extremely high average infrared power requirement, rising costs, and complexity of the laser systems.
As an alternative, resonant enhancement of optical pulses can be exploited instead of directly driving the OR process with complex high power amplifiers.
Until now, significant results have only been reported from active optical cavities, i.e., optical parametric oscillators (OPOs) and laser cavities.
For example, an Yb solid-state laser cavity was used for OR with a gallium phosphide (GaP) crystal at \SI{22}{\watt} circulating power to generate \SI{150}{\micro\watt} of THz at \SI{80}{\mega\hertz} repetition rate \cite{Hamrouni2021}.
At the end of 2023, milliwatt-level single-cycle THz pulse generation was demonstrated using a thin lithium niobate (LN) crystal of \SI{50}{\micro\meter} thickness placed inside a \SI{44.8}{\mega\hertz} repetition rate thin-disk oscillator \cite{Wang2023}.
Here the driving intracavity power was \SI{264}{\watt} and the extracted THz radiation exceeded \SI{1.3}{\milli\watt} with a \SI{3}{\tera\hertz} bandwidth.
Note that a thin LN plate offers distinct advantages over bulk LN crystals. 
Thin LN plates effectively reduce group velocity mismatch and thermal distortion, which commonly limit efficiency in bulk LN systems.
Hence, thin LN allows effective OR processes even at kilowatt-level driving powers.
Moreover, thin LN plates exhibit broad phase matching conditions, lower THz re-absorption, and reduced multiphoton absorption of the pump radiation.
An alternative approach to further increase the driving power levels involves the use of a passive enhancement cavity (EC). 
This method has the advantage of requiring less specialized know-how on mode-locked laser design and decoupling the THz generation to the mode-locking dynamics, which adds a degree of complexity. 
A standard watt-level laser with a pulse repetition rate in the range from \SI{40}{\mega\hertz} to several GHz can be coupled to an external EC that boosts the pulse energy by coherently summing multiple pulses, and, depending on the finesse of the cavity, passive gains of the order of thousands can be reached \cite{Pupeza2010, Carstens2014}.
However, cavity-enhanced THz generation has experienced only limited progress, in contrast to the success of ECs in frequency upconversion \cite{Pupeza2021, Kulpe2020, Zhang2024}.
To the best of our knowledge, only one attempt at using femtosecond EC for THz generation was made in 2008 by M. Theuer et al. \cite{Theuer2008}. This experiment was based on cavity-enhanced OR inside a bulk lithium niobate in Cherenkov-radiation type geometry. However, cavity losses constrained the maximum gain to $8.5$, corresponding to less than $\SI{7}{\watt}$ of circulating average power. Other attempts have been made recently with GaP crystals in collinear OR configuration \cite{Suerra23,Canella24};
nevertheless, the high multi-photon absorption and the strong temperature sensitivity of this material make its application for cavity-enhanced THz generation unpractical \cite{Hekmat2020}.
In fact, the buildup of significant power in a passive cavity relies on extremely low-loss nonlinear crystals.
In~\cite{Wang2023}, it was shown that this issue can be circumvented using thin lithium niobate plates.
\\
In this work, we implement the use of a thin lithium niobate plate with a thickness of \SI{50}{\micro\meter} for OR in a high-power enhancement resonator, driven by a commercial mode-locked Yb fiber laser.
In particular, the exceptionally low losses of our anti-reflection (AR) coated LN allows us to reach a passive gain of $240$, which corresponds to overcoming the kW-level intracavity power with an input Yb-fiber laser power of less than \SI{10}{\watt}.
With this configuration, we reached \SI{0.65}{\milli\watt} of measured THz average power at \SI{93}{\mega\hertz} repetition rate.
To the best of our knowledge, this is the first example of a high-gain, high-power EC used for single-pulse THz generation, and the first demonstration of OR guided at more than \SI{1}{\kilo\watt} average power.
Our conversion efficiency was limited mainly by the driving pulse duration of the laser system, offering a straightforward path to improve the result in future demonstrations. 
This configuration opens up unexplored scenarios for high repetition rate THz sources, because the EC can efficiently operate in a wide spectral region from visible to mid-infrared, and our approach greatly reduces the power requirements and complexity of ultrafast lasers to drive OR processes for high-repetition rate THz generation.

\section{\label{sec:exper}Experimental setup and measurements}
\begin{figure}
\centering
\includegraphics[width=8.5cm]{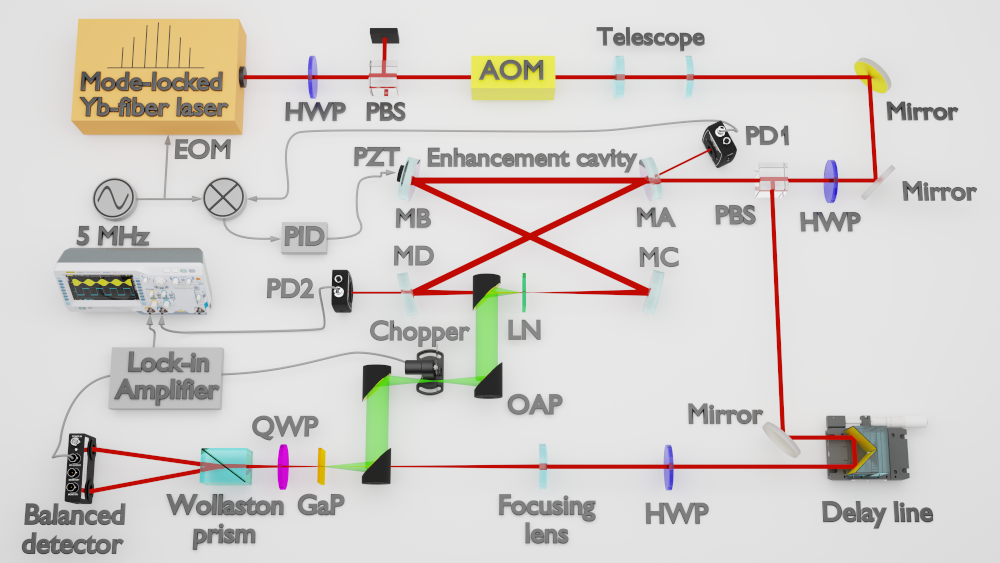}
\caption{Experimental setup.
HWP/QWP half/quarter-wave plate;
PBS polarizing beam splitter;
EOM/AOM electro/acousto-optic modulator;
PD$_\mathrm{i}$ photodetector;
M$_\mathrm{i}$ cavity mirror;
PZT piezo actuator;
LN lithium niobate plate;
OAP off-axis parabolic mirror.
}
\label{fig:setup}
\end{figure}

\begin{figure}
\centering
\includegraphics[width=4.25cm]{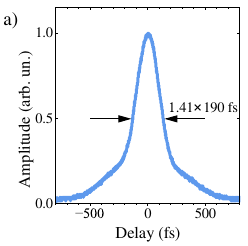}
\includegraphics[width=4.25cm]{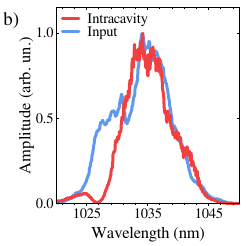}
\caption{SH intensity autocorrelation of the laser (a) and optical spectrum of the laser and of the intracavity beam (b).}
\label{fig:beamparam}
\end{figure}

\begin{figure}
\centering
\includegraphics[width=8.5cm]{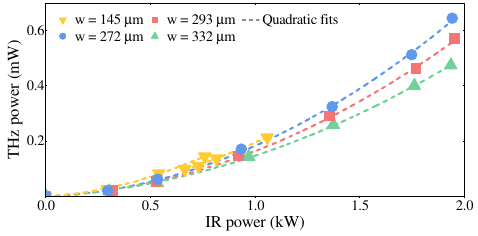}
\caption{Measured THz power as a function of IR pump power, for different IR beam radii (symbols).
Quadratic fits are plotted for each experimental curve (dashed lines).}
\label{fig:power}
\end{figure}

\begin{figure}
\centering
\includegraphics[width=4.25cm]{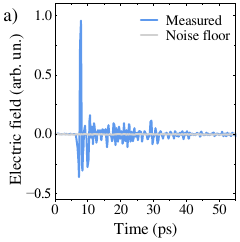}
\includegraphics[width=4.25cm]{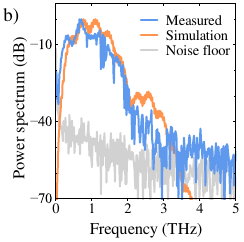}
\caption{(a) THz electric field measured by EOS (blue) and noise floor (gray).
(b) Calculated (blue) and simulated (orange) THz power spectrum, together with noise floor (gray).}
\label{fig:EOS}
\end{figure}

The experimental setup is reported in Fig.~\ref{fig:setup}.
We used the same LN crystal of Ref.~\cite{Wang2023}, where its complete characterization can be found.
It has a thickness of \SI{50}{\micro\meter} that helps to avoid high group velocity mismatch in a collinear geometry between the IR pump pulses and the generated THz radiation, simultaneously maintaining a high conversion efficiency.
The dimensions of the plate are $\SI{17}{\milli\meter}\times\SI{15}{\milli\meter}$, and it is oriented so that an input polarization along the extraordinary axis encounters the maximum nonlinear coefficient $\mathrm{d_{33}}$ of the crystal.
The plate is also AR-coated for a pump wavelength of \SI{1030}{\nano\meter} with a five-layer dielectric stack.
The laser source is a commercially available Yb-doped mode-locked fiber laser, that generates IR pulses with a FWHM length of \SI{250}{\femto\second} at \SI{92.9}{\mega\hertz}, with a FWHM spectral bandwidth of \SI{10.4}{\nano\meter} centered at \SI{1035}{\nano\meter}, and with a maximum output average power of \SI{9}{\watt} (model Orange, Menlo System).
The EC consists of a four-mirror bow-tie cavity with a free spectral range of \SI{92.9}{\mega\hertz}.
The reflectivity of the input plane mirror $\mathrm{MA}$ is \SI{99.2}{\percent}, while the one of the other mirrors exceeds $\SI{99.99}{\percent}$.
$\mathrm{MB}$ is plane, while $\mathrm{MC}$ and $\mathrm{MD}$ have a radius of curvature of \SI{750}{\milli\meter}.
The EC has a nominal finesse of $750$, and it is in a near-confocal configuration, allowing for a reduction in the spot size in the focal point, but at the same time maintaining negligible transversal astigmatism of the beam ($w_y/w_x > \SI{97}{\percent}$, with $w_{x,y}$ horizontal and vertical beam radii, respectively).
The lithium niobate plate is placed in the focusing branch so that the IR pump beam radius can be chosen between \SI{145}{\micro\meter} and \SI{600}{\micro\meter}.
Thanks to the extremely low thickness of the LN, the global intracavity group delay dispersion (GDD) is low.
In particular, the LN plate contributes with \SI{15.2}{\femto\second\squared}, while the cavity (dielectric mirrors and air) introduces a GDD of about \SI{80}{\femto\second\squared}, leading to a total GDD $<\SI{100}{\femto\second\squared}$.
Therefore, the intracavity pulse duration is only slightly affected, without any spectral clipping: with a finesse of $750$, the effect of this GDD is negligible.
The laser power directed to the EC can be adjusted between \SI{100}{\milli\watt} and \SI{9}{\watt} with a tunable attenuator, consisting of a half-wave plate and a polarizing beam splitter (PBS).
An acousto-optic modulator (AOM) is used to amplitude-modulate the laser beam and measure the EC frequency response, from which precise values of its finesse and gain can be calculated using the technique described in Ref.~\cite{Galzerano2020}.
Finally, another PBS extracts a small amount ($\approx\SI{500}{\milli\watt}$) of power for electro-optical sampling (EOS), and also guarantees that the beam coupled to the cavity is purely p-polarized.
Intracavity power is monitored with a photodiode (Thorlabs PDA36A) on the transmission of mirror $\mathrm{MD}$, and the experimental value of the gain ($328$ without the LN plate) has been used to find the conversion coefficient from voltage to intracavity power.
Notice that the maximum EC gain is reached when the laser and the cavity modes are perfectly matched \cite{Canella2022}, and this configuration also ensures minimization of the temporal elongation of pulses inside the EC \cite{Jones2002, Holzberger2015}.
A precise tuning of the laser spectral modes can be obtained by acting on its carrier-envelope offset (CEO), whose tuning is enabled by a motorized prism-pair inside our laser cavity.
Frequency locking of the EC to the laser is performed with a standard Pound-Drever-Hall (PDH) technique \cite{Drever1983} by frequency-modulating the laser pulses with its internal electro-optical modulator (EOM) at \SI{5}{\mega\hertz}.
The PDH error signal is processed by a PID and applied to a piezoelectric actuator attached to mirror $\mathrm{B}$, with a loop control bandwidth of nearly \SI{1}{\kilo\hertz}.
In these measurements, carrier-envelope offset stabilization was not required due to the minimal drift observed ($<\SI{100}{\kilo\hertz\per\minute}$) relative to the measurement duration ($\approx \SI{6}{\minute}$).
However, active CEO stabilization becomes essential for spectroscopic applications demanding extended stability over time.
An additional feedback loop would be required to achieve this, as demonstrated by Jones et al. \cite{Jones2001}, using components akin to those in Pound-Drever-Hall control.
The primary advantage of this configuration is that it requires only the stabilization of the relative offset between the laser and cavity combs, eliminating the need for an absolute reference or nonlinear optical components such as f-2f interferometers \cite{Telle1999}.
THz radiation is collected by a \SI{50.8}{\milli\meter}-focal length off-axis parabolic (OAP) mirror with a \SI{6}{\milli\meter}-diameter hole along its focus axis, placed inside the EC, and collimated to a \SI{101.6}{\milli\meter}-focal length OAP.
Then the THz is focused on a chopper, used for power and EOS measurements
A third \SI{101.6}{\milli\meter}-focal length OAP collimates THz radiation again and sends it to a last \SI{50.8}{\milli\meter}-focal length OAP focusing on the power meter or the GaP crystal for EOS.
This last OAP has a \SI{3}{\milli\meter} diameter hole to allow the IR probe to overlap with THz for the EOS measurement.
All OAPs have an external diameter of \SI{50.8}{\milli\meter} and protected gold coating.
Power measurements have been performed with a THz power meter (Ophir RM9-THz) chopping at \SI{18}{\hertz}, and placing two calibrated black sheets (THz transmission \SI{20}{\percent}) along the THz path to filter out all residual power both at \SI{1030}{\nano\meter} wavelength, and at the second harmonic wavelength of \SI{515}{\nano\meter}.
The background floor of the THz power meter was $<\SI{1}{\micro\watt}$.
\\
Concerning intracavity pulse parameters, Fig.~\ref{fig:beamparam} shows the second harmonic (SH) intensity autocorrelation of the laser beam at cavity input (a), together with the spectra of the laser beam at the cavity input and intracavity (b).
Notice that the spectrum of the intracavity pulse is not directly accessible, thus we equivalently measured the spectrum of the transmitted beam.
The FWHM of the autocorrelation trace is \SI{270}{\femto\second}, indicating a Gaussian pulse with a FWHM of \SI{190}{\femto\second}.
On the other hand, the spectrum has a \SI{10.4}{\nano\meter} FWHM, yielding a \SI{150}{\femto\second} FWHM transform-limited Gaussian pulse, thus the effective time-bandwidth product (TBP) is $0.555$.
An estimation of the EC pulse length has been performed from the transmitted spectrum (\SI{7.9}{\nano\meter} FWHM) by maintaining the same TBP of the input beam (EC nonlinear effects are negligible), resulting in \SI{250}{\femto\second} FWHM.
The observed spectral reduction is due to a measured laser spectrum variation along the beam profile.
The single spatial $\mathrm{TEM_{00}}$ mode profile coupled into the EC therefore results in a slightly narrower intracavity spectral bandwidth.
Despite all these limitations due to the laser properties, promising results have been obtained using a simple commercially available laser system, demonstrating the potential of the EC approach.
\\
Figure \ref{fig:power} shows THz power measurement as a function of intracavity IR power for different radii of the IR beam incident on the LN crystal.
The trend is quadratic for almost every beam dimension, without evident saturation behavior.
We measured a maximum THz power of \SI{0.65}{\milli\watt} with an intracavity power of \SI{1.9}{\kilo\watt}, corresponding to an IR peak intensity of \SI{50}{\giga\watt\per\centi\meter\squared}.
The maximum THz power is generated for $w=\SI{272}{\micro\meter}$.
On the other hand, we noticed a drop in THz power when the IR beam radius was \SI{145}{\micro\meter} and peak intensity exceeded \SI{70}{\giga\watt\per\centi\meter\squared} (see Fig.~\ref{fig:power}, yellow symbols).
At this intensity, we observed the generation of IR at \SI{1100}{\nano\meter} and its SH at \SI{550}{\nano\meter}.
The direction of propagation of \SI{1100}{\nano\meter} and \SI{550}{\nano\meter} light was also at a different angle with respect to the pump beam.
Since the IR redshift corresponds to well-known phonon resonance of LN at \SI{18.97}{\tera\hertz} \cite{Gorelik2017}, our hypothesis is that an interaction with phonons occurs.
This phenomenon could constitute a limiting factor for the maximum THz power, and it will be better investigated in a forthcoming work.
In the condition of maximum THz power (\SI{0.65}{\milli\watt}), we reached an efficiency in THz generation of \SI{3.4e-7}{} with respect to intracavity power (\SI{1.9}{\kilo\watt}), or \SI{0.8e-4}{} concerning the input power (\SI{8}{\watt}).
Further considerations could be addressed to estimating the real THz power generated with the LN plate, correcting the measured value for water absorption and spectral filtering of OAPs due to THz diffraction.
Atmospheric water vapor absorption along the \SI{66}{\centi\meter} path between LN and THz detector is \SI{9.0}{\percent}.
The spectral filtering of the first OAP contributes to a power loss of \SI{57.9}{\percent} in the condition of maximum power generation in our setup (THz beam radius of \SI{192}{\micro\meter}).
Overall, we estimate a total power loss of \SI{61.7}{\percent}, leading to a generated THz power of \SI{1.7}{\milli\watt}, corresponding to a THz generation efficiency of \SI{2e-4}{}, which comes closer to other values reported in literature.
Notice that OAP reflectivity has been considered \SI{100}{\percent}.
\\
A small fraction of the laser power is used for EOS of THz pulses (see Fig.~\ref{fig:setup}) with a conventional balanced detection technique, employing a \SI{1}{\milli\meter} AR-coated GaP crystal, a quarter-waveplate, a Wollaston prism, and a balanced detector.
A motorized translation stage was placed along the probe beamline and used to scan the time delay between the probe IR and the THz pulses in a range of \SI{70}{\pico\second} and with a measurement time of \SI{350}{\second}.
The chopper was set at a frequency of \SI{425}{\hertz} for low noise detection of EOS signal by means of a lock-in amplifier set (Stanford Research Systems SR530), with an integration time of \SI{1}{\second}.
The recorded THz electric field as a function of time is shown in Fig.~\ref{fig:EOS}(a) in the condition of maximum THz power generation, together with the calculated THz power spectrum presented in Fig.~\ref{fig:EOS}(b).
We generate a single-cycle THz pulse with a carrier frequency of \SI{1}{\tera\hertz} and a spectrum up to \SI{2}{\tera\hertz} and a dynamic range of \SI{50}{\decibel} above the noise floor (acquired blocking the THz beam).
The moderate bandwidth is consistent with the pulse length of the driving laser.
In the THz spectrum, two etalon modulations are visible, with frequencies equal to \SI{47}{\giga\hertz} and \SI{420}{\giga\hertz}, corresponding to the thickness of
the GaP used for EOS and to the thin LN, respectively.
Fig.~\ref{fig:EOS}(b) shows also a comparison with the simulated spectrum (orange), calculated by solving the coupled wave equations for OR following the procedure reported in Ref.~\cite{Hattori07}, where we have included the response of the \SI{1}{\milli\meter} GaP crystal, the effect of the OAPs spectral filtering \cite{Faure2004}, and echoes from both of the LN and GaP reflections.
We found a good agreement in the low-frequency part of the spectrum up to nearly \SI{2.0}{\tera\hertz}, and the expected thickness of the crystal is well represented by the dips.
A small deviation can be noticed at higher frequencies, most likely due to a lack of precise literature data on the THz refractive index of lithium niobate beyond \SI{2}{\tera\hertz} at room temperature \cite{Wu2015,Unferdorben2015}, and to its dependence on temperature changes induced by such a high intracavity power level.
We remark that the main limitation of our setup derives from the commercial laser's features, most prominently its long pulse duration.
However, we demonstrated promising results of cavity-enhanced THz overcoming the laser limitations.
Remarkably, we also show that lithium niobate can operate damage-free and with reasonable conversion efficiency at \SI{1.9}{\kilo\watt} of average power.

\section{\label{sec:conclu}Conclusions}
In summary, we propose and demonstrate a simple, and cost-effective approach for high-repetition-rate, high-power THz generation using a thin lithium niobate plate in an external enhancement cavity, seeded by a \SI{8}{\watt} average power commercial Yb-fiber mode-locked laser.
To the best of our knowledge, this is the first example of OR with a crystal pumped with a \SI{1.9}{\kilo\watt} average power at \SI{92.9}{\mega\hertz} repetition rate. 
We measured \SI{0.65}{\milli\watt} single-cycle THz pulses, with a spectrum of \SI{2}{\tera\hertz} width.
We believe that the main limitation of our setup is the long duration of the pulses produced by the commercial laser.
This relatively compact and cost-effective approach can provide a new class of high-repetition-rate THz sources by overcoming this issue and improving the conversion efficiency to $10^{-6}$.
\section*{Author Declaration}
\subsection*{Conflict of Interest}
The authors have no conflicts to disclose.

\subsection*{Preprint declaration}
The following article has been submitted to AIP Publishing Group.

\subsection*{Copyright}
Copyright 2025 Authors. This article is distributed under a Creative Commons Attribution-NonCommercial-NoDerivs 4.0 International (CC BY-NC-ND) License.

\section*{Data Availability Statement}
The data that support the findings of this study are available from the corresponding author upon reasonable request.

\begin{acknowledgments}
This work has been supported by: 
INFN Gruppo V, project ETHIOPIA;
European Union’s NextGenerationEU Program with the I-PHOQS Infrastructure [IR0000016, ID D2B8D520, CUP B53C22001750006];
Ministerium für Kultur und Wissenschaft des Landes Nordrhein-Westfalen (terahertz.NRW);
Deutsche Forschungsgemeinschaft (287022738 TRR 196, 390677874, RESOLV);
HORIZON EUROPE European Research Council (805202).
\\
The authors thank Dr. Federica Bianco and Prof. Alessandro Tredicucci for their support in the calibration of THz sensors.
\end{acknowledgments}

\bibliography{bibliography}

\end{document}